\documentclass[acmsmall,screen]{acmart}

\usepackage{amsmath}
\usepackage{amsthm}

\usepackage[utf8]{inputenc} 
\usepackage[T1]{fontenc}    
\usepackage{hyperref}       
\usepackage{url}            
\usepackage{booktabs}       
\usepackage{makecell}
\usepackage{amsfonts}       
\usepackage{nicefrac}       
\usepackage{microtype}      
\usepackage{xcolor}         

\usepackage{graphicx}
\usepackage{lipsum}  

\usepackage{listings}
\usepackage{algorithm}
\usepackage{algpseudocode}

\usepackage{caption}
\usepackage{subcaption}

\usepackage{pifont}
\usepackage{multirow}

\usepackage{tablefootnote} 

\usepackage{caption}
\definecolor{codegreen}{rgb}{0,0.6,0}
\definecolor{codegray}{rgb}{0.5,0.5,0.5}
\definecolor{codepurple}{rgb}{0.58,0,0.82}
\definecolor{backcolour}{rgb}{0.95,0.95,0.92}

\lstdefinestyle{mystyle}{
    commentstyle=\color{codegreen},
    keywordstyle=\bf\ttfamily\color[rgb]{0,.3,.7},
    numberstyle=\color{codegray},
    stringstyle=\color{red}\ttfamily,
    basicstyle=\ttfamily,
    breakatwhitespace=false,         
    breaklines=true,                 
    captionpos=b,                    
    keepspaces=true,                 
    numbers=left,                    
    numbersep=5pt,                  
    showspaces=false,                
    showstringspaces=false,
    showtabs=false,                  
    tabsize=2
}

\usepackage{color}
\definecolor{lightgray}{rgb}{.9,.9,.9}
\definecolor{darkgray}{rgb}{.4,.4,.4}
\definecolor{purple}{rgb}{0.65, 0.12, 0.82}
\lstdefinelanguage{JavaScript}{
  keywords={break, case, catch, continue, debugger, default, delete, do, else, false, finally, for, function, if, in, instanceof, new, null, return, switch, this, throw, true, try, typeof, var, void, while, with},
  morecomment=[l]{//},
  morecomment=[s]{/*}{*/},
  morestring=[b]',
  morestring=[b]",
  ndkeywords={class, export, boolean, throw, implements, import, this},
  sensitive=true
}

\lstdefinelanguage{julia}%
  {morekeywords={abstract,break,case,catch,const,continue,do,else,elseif,%
      end,export,false,for,function,immutable,import,importall,if,in,%
      macro,module,otherwise,quote,return,switch,true,try,type,typealias,%
      using,while},%
   sensitive=true,%
   alsoother={\$},%
   morecomment=[l]\#,%
   morecomment=[n]{\#=}{=\#},%
   morestring=[s]{"}{"},%
   morestring=[m]{'}{'},%
}[keywords,comments,strings]%

\lstdefinelanguage{scheme}{
  morekeywords=[1]{define, define-syntax, define-macro, lambda, define-stream, stream-lambda},
  morekeywords=[2]{begin, call-with-current-continuation, call/cc,
    call-with-input-file, call-with-output-file, case, cond,
    do, else, for-each, if,
    let*, let, let-syntax, letrec, letrec-syntax,
    let-values, let*-values,
    and, or, not, delay, force,
    quasiquote, quote, unquote, unquote-splicing,
    map, fold, syntax, syntax-rules, eval, environment, query },
  morekeywords=[3]{import, export},
  alsodigit=!\$\%&*+-./:<=>?@^_~,
  sensitive=true,
  morecomment=[l]{;},
  morecomment=[s]{\#|}{|\#},
  morestring=[b]",
  upquote=true,
  breaklines=true,
  breakatwhitespace=true,
  literate=*{`}{{`}}{1},
  showstringspaces=false
}

\newlength{\myMheight}
\usepackage{array}
\newcolumntype{H}{>{\setbox0=\hbox\bgroup}c<{\egroup}@{}}

\newtheorem{definition}{Definition}
\newtheorem{theorem}{Theorem}

\author{Hebi Li}
\email{lihebi.com@gmail.com}
\affiliation{%
  \institution{CodePod Inc.}
  \city{Santa Clara}
  \state{California}
  \country{USA}
}

\author{Forrest Sheng Bao}
\email{forrest.bao@gmail.com}
\affiliation{%
  \institution{CodePod Inc.}
  \city{Santa Clara}
  \state{California}
  \country{USA}
}

\author{Qi Xiao}
\email{xiaoqiqiduo@gmail.com}
\affiliation{%
 \institution{LinkedIn}
 \city{Sunnyvale}
 \state{California}
 \country{USA}
}

\author{Jin Tian}
\email{Jin.Tian@mbzuai.ac.ae}
\affiliation{%
  \institution{MBZUAI}
  \city{Dubai}
  \state{United Arab Emirates}
  \country{UAE}
}

\begin{document}

\title[CodePod: A Language-Agnostic Hierarchical Scoping System for Interactive Development]{CodePod: A Language-Agnostic Hierarchical Scoping System for Interactive Development}

\begin{abstract}
  Interactive development environments like Jupyter Notebooks enable incremental coding through cells with immediate feedback, but their linear structure and global namespace limit scalability for large software projects. We present CodePod, a hierarchical extension of Jupyter that introduces a novel scoped execution model with formal semantics. Our key contribution is a language-agnostic runtime system that performs source-level transformations to implement hierarchical scoping rules, enabling true incremental evaluation across nested modules without requiring language-specific kernel modifications. We formalize the scoping semantics as a mathematical framework with precise visibility relations and prove key properties including uniqueness of symbol resolution and correctness of the resolution algorithm. A qualitative user study with seven senior developers demonstrates that CodePod enables significant improvements in project scalability compared to Jupyter, with notable reductions in navigation effort. We validate the system's effectiveness on large-scale projects with thousands of lines of code, demonstrating its applicability beyond traditional notebook boundaries. Our tool is available at \url{https://codepod.io}.
  
\end{abstract}

\maketitle


\section{Introduction}

Interactive development environments have become essential tools for modern software development, particularly in domains requiring rapid prototyping and exploratory programming. Jupyter Notebooks~\cite{jupyter} exemplify this paradigm, offering a browser-based environment that integrates code, text, and execution results in a literate programming model. By supporting incremental evaluation through a Read-Eval-Print-Loop (REPL) model~\cite{mccarthy1965lisp,van2020principled}, Jupyter has achieved widespread adoption in data science, machine learning, and education~\cite{wang2020better,randles2017using,perkel2018jupyter,pimentel2019large}. The 2024 GitHub Octoverse survey reports over 1.5 million GitHub repositories with Jupyter Notebooks, marking a 92\% year-over-year increase~\cite{github_octoverse_2024}.

Despite its popularity, Jupyter's design exhibits fundamental limitations that hinder its applicability to large-scale software development~\cite{head2019managing}. The primary architectural constraints stem from two design decisions: (1) a linear cell arrangement that lacks hierarchical organization, making dependency management and navigation increasingly complex as projects grow; and (2) a global namespace model that precludes proper modularity, leading to naming conflicts and inhibiting structured development practices. These limitations reflect a broader tension between the benefits of interactive development and the organizational requirements of large-scale software engineering.

To address these challenges, we present CodePod, a hierarchical extension of Jupyter that introduces a novel scoped execution model with formal semantics. As illustrated in Figure~\ref{fig:motivating_pitch}, CodePod transforms Jupyter's linear structure into a nested two-dimensional canvas that supports:

\begin{figure}[ht]
    \centering
    \includegraphics[width=0.85\textwidth]{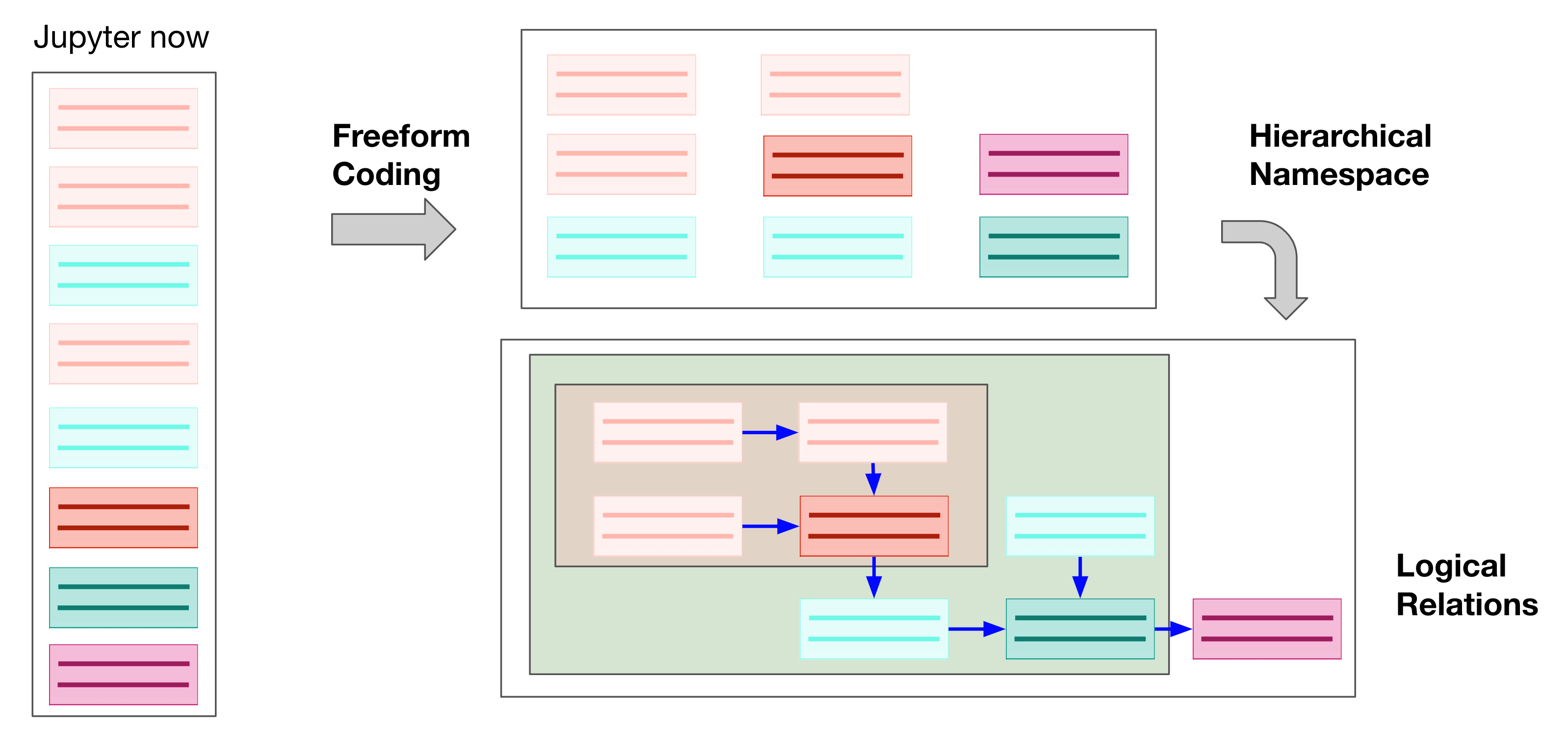}
    \caption{From Jupyter's linear structure to a nested 2D canvas with hierarchical scoping}
    \label{fig:motivating_pitch}
\end{figure}

\begin{itemize}
    \item \textbf{Spatial Organization:} Code blocks can be freely positioned on a nested two-dimensional canvas, enabling intuitive spatial organization that leverages human spatial memory for complex project navigation.
    \item \textbf{Hierarchical Scoping:} Related code blocks can be grouped into \textit{scopes} that encapsulate variables and functions, preventing naming conflicts and enabling structured, hierarchical organization with explicit API boundaries.
    \item \textbf{Dependency Visualization:} Automatic analysis and visualization of \textit{definition-use} relationships provides immediate insight into code dependencies and logical structure.
\end{itemize}

Our work makes two key technical contributions to the programming languages and interactive development communities. First, we develop a set of semantic scoping rules governing the interaction of functions and modules within our hierarchical framework. We derive a mathematical framework that precisely defines visibility and accessibility relationships between code elements in a hierarchical structure. This framework extends traditional lexical scoping concepts to support nested modules with explicit export mechanisms, providing a rigorous foundation for reasoning about program structure and dependencies.

Second, we introduce a language-agnostic runtime system that supports these semantic scoping rules. We introduce a novel approach to implementing hierarchical scoping through source-level transformations rather than kernel modifications. This design enables true incremental evaluation across nested modules while maintaining compatibility with existing language kernels, supporting multiple programming languages without requiring language-specific implementations.


Our evaluation demonstrates the effectiveness of this approach through both theoretical analysis and empirical validation. A qualitative user study with seven senior developers shows that CodePod enables 1.33x to 4x improvement in project scalability compared to Jupyter, with significant reductions in navigation effort. We further validate the system's effectiveness on large-scale projects with thousands of lines of code, demonstrating its applicability beyond traditional notebook boundaries.

The remainder of this paper is organized as follows: Section~\ref{sec:motivating} motivates the problem through concrete examples; Section~\ref{sec:approach} presents our formal semantics and runtime system; Section~\ref{sec:eval} evaluates our approach through user studies and large-scale validation; and Section~\ref{sec:related} discusses related work.

\section{The Scalability Challenge of Jupyter Notebooks}
\label{sec:motivating}

\subsection{Organizational Limitations within Notebooks}

Interactive development environments like Jupyter Notebooks have become ubiquitous in data science and machine learning workflows, yet their fundamental design exhibits inherent limitations that impede scalability. We illustrate these limitations through a concrete example: a neural network classifier for the Fashion-MNIST dataset implemented in a traditional Jupyter notebook.

\begin{figure}[ht]
    \centering
    \includegraphics[width=0.8\textwidth]{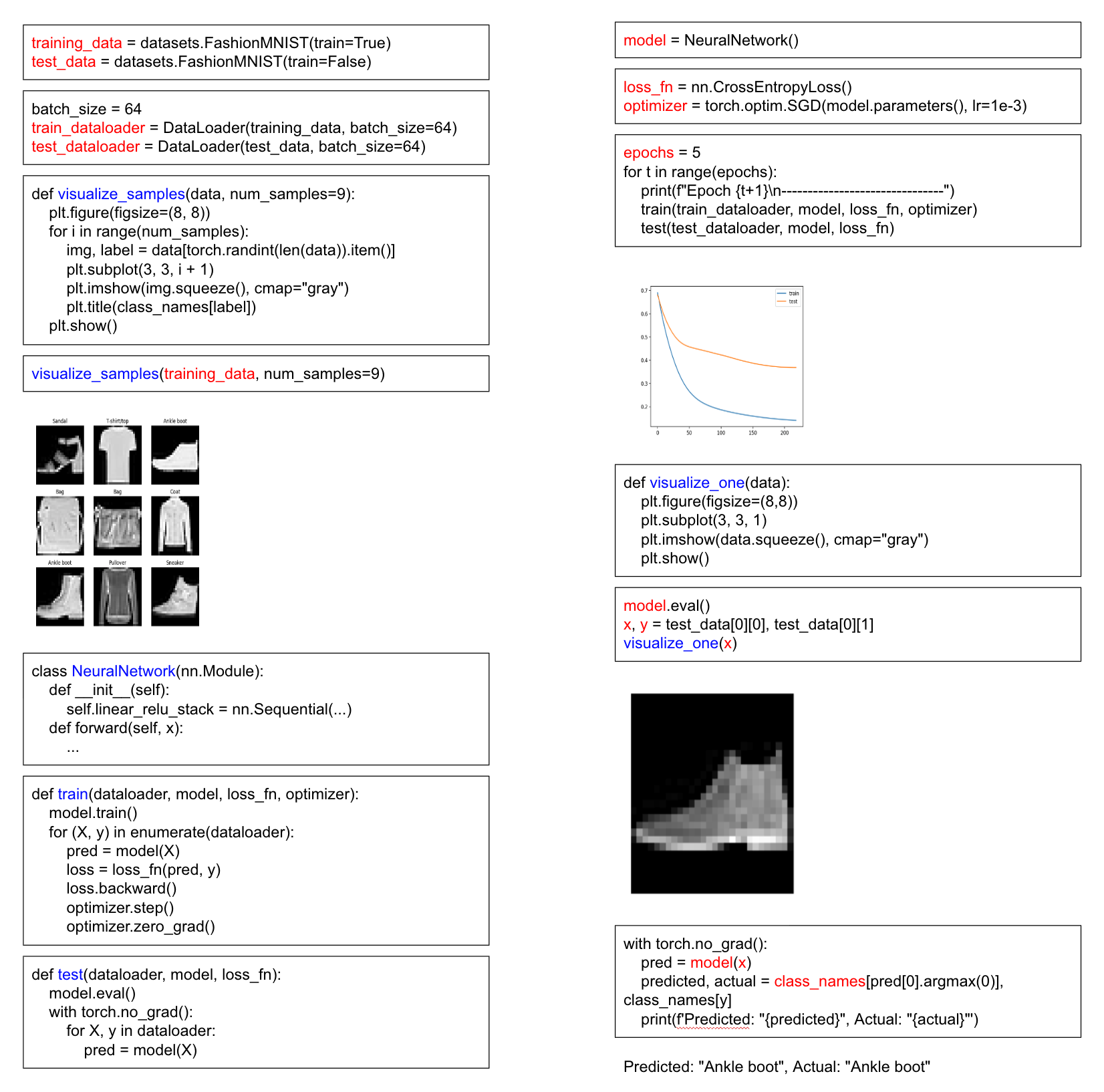}
    \caption{A Fashion-MNIST image classifier implemented in Jupyter Notebook, demonstrating the linear organization that becomes problematic as project complexity increases. The same project will be shown reorganized in CodePod in Figure~\ref{fig:tool-interface}.}
    \label{fig:mnist-jupyter}
\end{figure}

The notebook depicted in Figure~\ref{fig:mnist-jupyter} exemplifies a typical machine learning workflow comprising three logical components: data preprocessing, model architecture definition and training, and model evaluation. While functionally complete, this implementation reveals two fundamental architectural constraints inherent to Jupyter's design.

First, the linear cell arrangement obscures the dependency relationships between code blocks. The data preprocessing functions at the top of the notebook are conceptually distant from the evaluation code at the bottom, despite their logical coupling. This spatial separation makes it difficult to understand the data flow and function dependencies, particularly as the number of code cells grows. The linear structure forces developers to mentally reconstruct the program's logical organization from its physical layout, creating cognitive overhead that scales poorly with project size.

Second, Jupyter's global namespace model precludes proper encapsulation and modularity. All functions, variables, and classes exist in a single namespace, creating the potential for naming conflicts and unintended side effects. For instance, a utility function defined in the data preprocessing section might inadvertently shadow or overwrite a function in the model definition section. This lack of scoping mechanisms makes it difficult to maintain clean interfaces between different components of the system.

\begin{figure}[ht]
    \centering
    \includegraphics[width=0.9\linewidth]{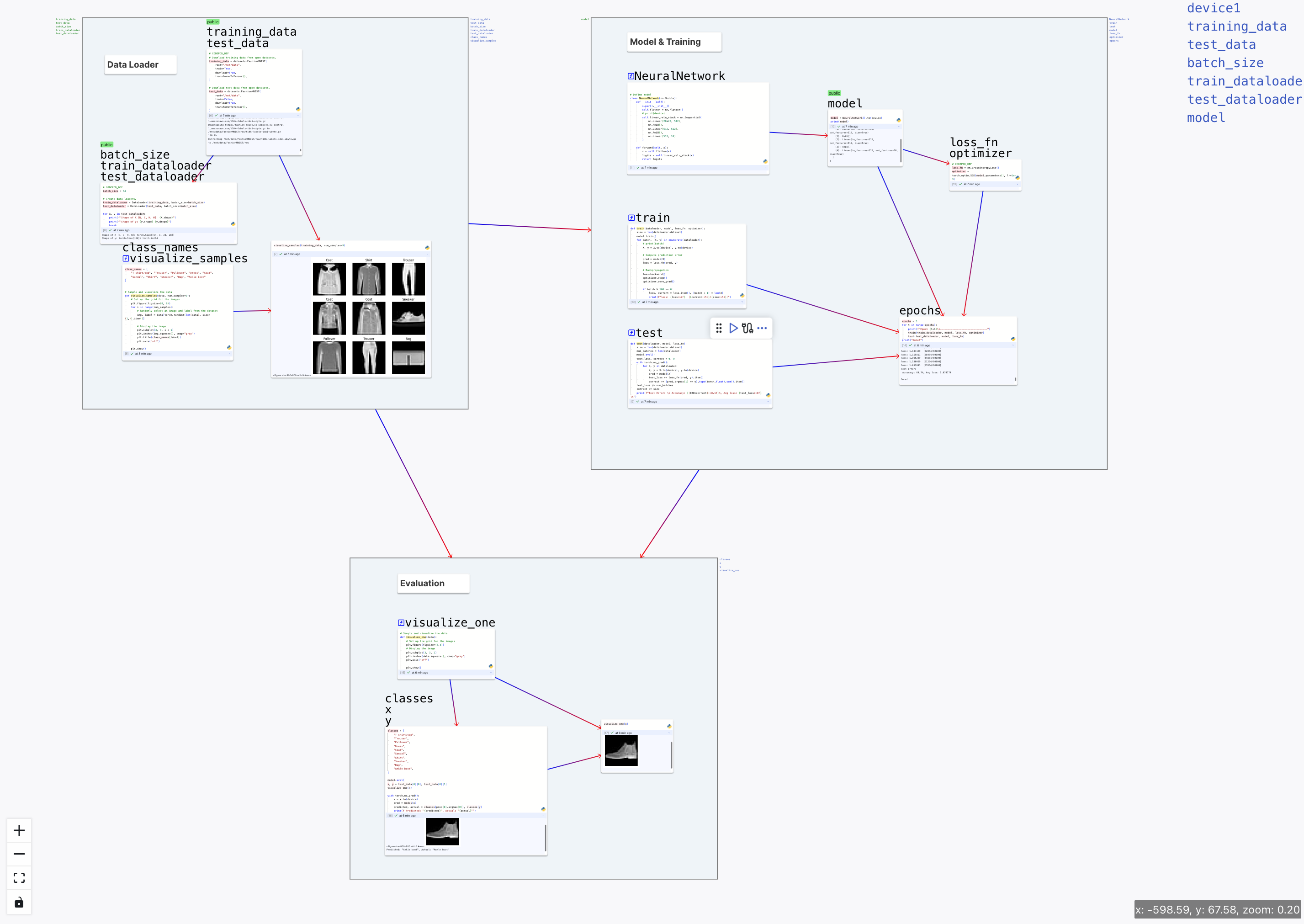}
    \caption{The same Fashion-MNIST classifier reorganized in CodePod, demonstrating hierarchical scoping and spatial organization. Code blocks are grouped into semantic scopes with explicit dependency visualization.}
    \label{fig:tool-interface}
\end{figure}

CodePod addresses these limitations through hierarchical organization and scoped execution, as illustrated in Figure~\ref{fig:tool-interface}. The code blocks are organized into semantically meaningful scopes that encapsulate related functionality. Each scope maintains its own namespace, preventing naming conflicts while enabling controlled sharing through explicit export mechanisms. The dependency relationships between code blocks are automatically analyzed and visualized, providing immediate insight into the program's structure and data flow.

\subsection{Scalability Beyond Single Notebooks}

The limitations of traditional Jupyter notebooks become particularly acute when considering large-scale software development projects. While Jupyter excels at exploratory programming and rapid prototyping, its design assumptions break down when applied to projects involving hundreds or thousands of functions across multiple modules.

Traditional Jupyter notebooks are optimized for data analysis workflows typically involving dozens of code cells. However, production software systems often comprise complex hierarchies of functions, classes, and modules with intricate interdependencies that cannot be effectively managed in a linear notebook format. The lack of hierarchical organization makes it difficult to understand the system's architecture, locate specific functionality, and maintain clean interfaces between components.

CodePod's hierarchical design addresses these limitations by providing a unified development environment that scales from individual notebooks to entire codebases. The scoped execution model naturally extends to large-scale projects, where entire files and modules can be represented as scopes containing their constituent functions and classes. This hierarchical organization preserves the interactive, cell-based development experience that makes notebooks valuable while providing the structural benefits typically associated with traditional integrated development environments.


This scalability is crucial for bridging the gap between exploratory development and production software engineering. CodePod enables developers to begin with rapid prototyping in a notebook-like environment and seamlessly scale their development process as the project grows, without losing the interactive development experience that makes notebooks valuable for exploration and experimentation.

\subsection{Challenges of Leveraging the File Systems}

Existing approaches attempt to address Jupyter's scalability limitations by integrating with traditional file systems and module systems. Jupyter's \texttt{\%run} magic command~\cite{ipython_magic_run} enables execution of Python files within notebook sessions, while tools like nbimporter~\cite{nbimporter_github} allow importing notebooks as modules. Additionally, Python's module system can be leveraged through \texttt{import} statements with autoreload functionality~\cite{ipython_autoreload}.

However, these approaches exhibit fundamental limitations that make them unsuitable for scalable interactive development:

\begin{itemize}
    \item \textbf{Lack of true incremental evaluation:} Different notebooks run on separate kernels (distinct processes) that do not share memory. When developing notebooks A and B interactively, importing B into A only creates a copy of B's functions and data in A's kernel, breaking the interactive development model. Notebook A cannot share states with notebook B, making it impossible to maintain a unified development environment.

    \item \textbf{Non-trivial and error-prone re-importing:} When imported functions are updated, they must be re-imported into the notebook session. This becomes complex in larger projects with interdependencies. For example, if notebook A imports function b() from module B, which depends on function c() from module C, and function c() changes, one must reload module C before reloading module B. The order of re-importing is significant and non-trivial to handle correctly. Other languages have even less support for such re-importing in Jupyter sessions.

    \item \textbf{Potential naming conflicts and lack of proper module systems:} Jupyter's magic command \textit{\%run} does not feature a module system, and imported notebooks are evaluated in the global namespace. A function in an imported notebook may accidentally overwrite a function defined in the importing notebook. While it is possible to assign the same kernel to multiple notebooks, they then execute in the same global namespace, suffering from the namespace conflict problem.

    \item \textbf{Unnecessary overhead and side effects:} Notebooks naturally contain both production code and exploratory elements (data loading, intermediate results, and temporary computations). When importing notebooks as modules, all code is executed, including these exploratory elements, creating unnecessary overhead and potential side effects.

    \item \textbf{Reduced Jupyter interactivity:} Regular Python files lack the interactivity provided by Jupyter. Different notebooks execute on separate kernels, making it hard to share data across notebooks, such as loaded (potentially expensive) datasets or intermediate results in variables.

    \item \textbf{Lack of fine-grained modularity within notebooks:} These approaches provide no modularity within a single notebook, resulting in the messiness within notebooks that we identified earlier. There is no way to organize related code blocks into logical groups within a notebook.

    \item \textbf{Language-specific limitations:} All these approaches work only in Python but not other languages supported by Jupyter. It is non-trivial to implement these approaches in other languages given the language-specific runtime and module system.
\end{itemize}

In contrast, CodePod provides a unified solution that addresses these limitations through hierarchical nested scopes and a language-agnostic scoped runtime system. The system enables true incremental evaluation through lightweight symbol table transformations, eliminating the need for complex re-importing procedures. By rewriting symbol references with unique identifiers, CodePod ensures that all code sees consistent, up-to-date definitions without requiring manual synchronization. Furthermore, CodePod provides fine-grained modularity within notebooks through nested scopes on a 2D canvas, superior dependency visualization through automatic def-use analysis, and a consistent module system that is native to the Jupyter/REPL development model.

\section{Our Approach}
\label{sec:approach}

This section presents the design and implementation of CodePod, a hierarchical extension of Jupyter that addresses the scalability limitations of traditional interactive development environments. CodePod enhances Jupyter's interactive and literate programming model through three fundamental innovations: (1) a nested two-dimensional canvas enabling flexible spatial organization of code blocks, (2) hierarchical scopes that provide structured encapsulation and visibility control, and (3) automatic def-use analysis that visualizes code dependencies and relationships.

Our primary technical contributions consist of two main components: semantic scoping rules that define symbol visibility, and a language-agnostic runtime system that implements these semantics through source-level transformations. We first present the semantic scoping rules that govern function and module interactions within our hierarchical framework, followed by the runtime architecture that supports these hierarchical scopes without requiring language-specific kernel modifications.

\subsection{Terminology and Scope}

We adopt the following terminology throughout this paper. A \emph{pod} refers to an individual code block containing executable code, analogous to a Jupyter cell. A \emph{scope} represents a logical container that groups related pods and establishes a namespace boundary. Note that in the literature, there have been diverse definitions of what constitutes a module, e.g.,~\cite{flatt1998units,corwin2003mj,ichisugi2002difference}. In this paper, when we refer to "namespace" or "module," we specifically address the concept of symbol visibility, which includes functions, variables, and other identifiers. By namespace or module, we mean the encapsulation of code elements to manage their visibility and accessibility within different parts of a program. This encapsulation allows for the creation of public APIs that expose certain functionalities while keeping other parts of the code private and hidden from external access.

In this paper, we focus on top-level programming constructs commonly used in interactive notebooks: top-level functions, classes, and variables. We explicitly exclude language-specific constructs such as imported symbols, type annotations, and metaprogramming features (e.g., macros in Racket/Julia) from our scoping semantics.

\subsection{The Nested Two-Dimensional Canvas}

CodePod's fundamental architectural innovation is the transition from Jupyter's linear cell arrangement to a nested two-dimensional canvas that enables flexible spatial organization of code blocks. This design choice is motivated by three key insights about software development at scale.

First, software systems exhibit inherent hierarchical structure, composed of functions, classes, and modules that form directed acyclic graphs of dependencies and relationships. Traditional Jupyter notebooks force this inherently hierarchical content into a linear sequence, creating cognitive overhead as developers must mentally reconstruct the program's logical organization from its physical layout. This limitation becomes increasingly problematic as projects grow in size and complexity.

Second, hierarchical organization is essential for scaling beyond the limitations of linear notebooks. While Jupyter's linear structure suffices for small projects with dozens of cells, it becomes unwieldy when managing hundreds of code blocks with complex interdependencies. The nested 2D canvas provides the organizational benefits of traditional integrated development environments while preserving Jupyter's interactive, cell-based development model.

Third, the 2D canvas offers spatial flexibility while maintaining hierarchical structure. During our design process, we evaluated three alternative approaches for implementing hierarchical organization: traditional tree structures, mind maps, and nested 2D canvases. Tree structures and mind maps proved too rigid for interactive development—they constrain the placement of related code blocks and make it difficult to arrange code in ways that reflect logical workflows and developer mental models. The nested 2D canvas preserves the tree structure's organizational benefits while allowing code blocks to be positioned flexibly based on their relationships and the developer's conceptual model of the project.

The canvas implementation supports zooming and panning operations, enabling developers to view both detailed code and high-level project structure. Code blocks can be freely positioned, resized, and grouped into scopes that represent logical modules. This spatial organization leverages human spatial memory capabilities, making it easier to navigate complex codebases and understand relationships between different components. The design choice has proven valuable both in our own development experience and in our user studies, where participants consistently reported improved navigation and project comprehension compared to traditional linear notebooks.

\subsection{Semantic Scoping Rules}
\label{sec:canvas}

\begin{figure}[ht]
    \centering
    \includegraphics[width=0.95\textwidth]{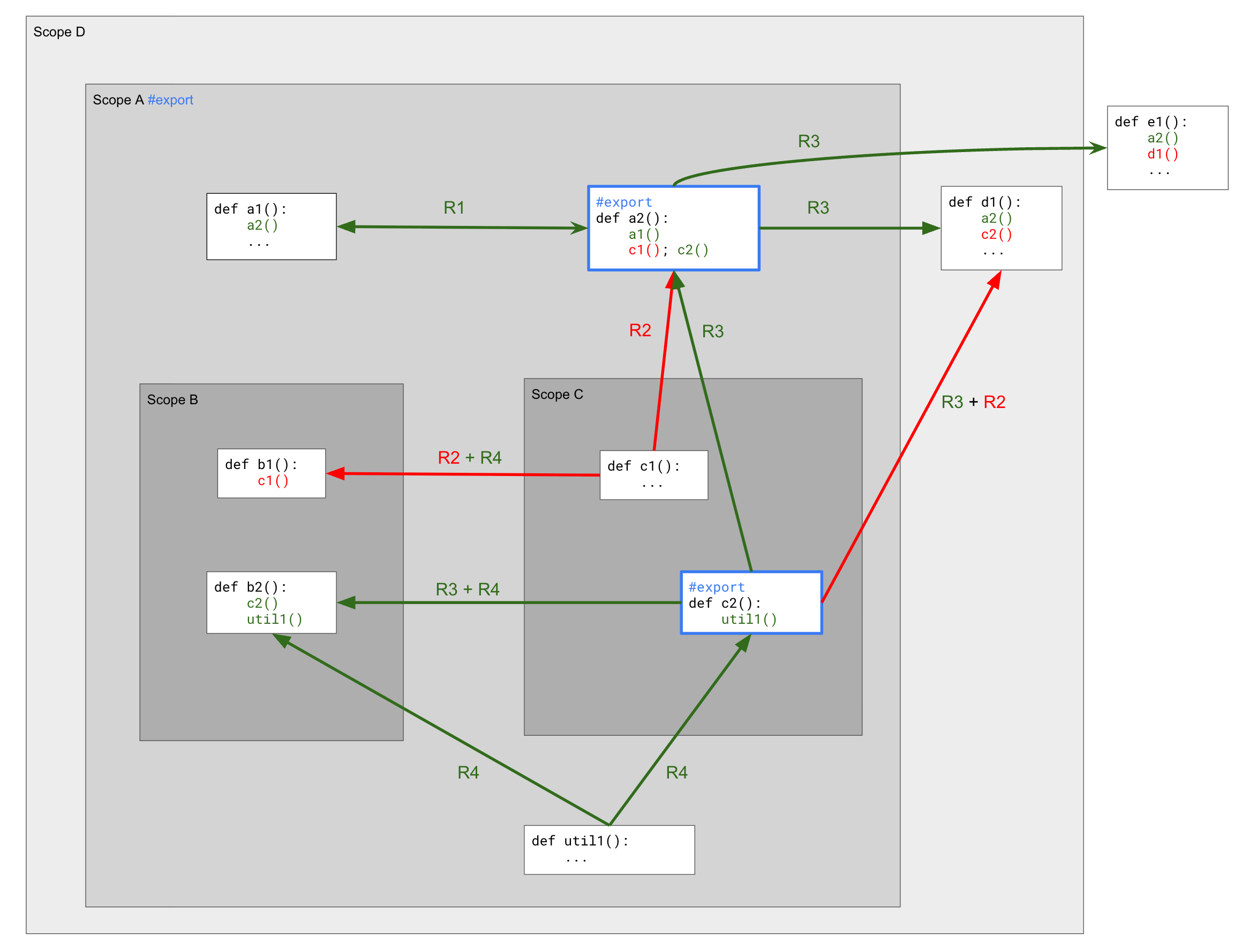}
    \caption{Scoping Rules. \textcolor{green}{R1} with a green arrow denotes that the function is visible along the arrow direction, under scoping rule R1. \textcolor{red}{R2} with a red arrow denotes that the function is non-visible along the arrow direction, under scoping rule R2. \textcolor{cyan}{\texttt{\#export}} marks a function as public API for the scope, making it visible to its parent scope.}
    \label{fig:semantic-rules}
\end{figure}

We establish a set of scoping rules that define how functions and modules interact within our hierarchical framework. These rules are critical in managing the visibility and accessibility of functions across different scopes. Figure~\ref{fig:semantic-rules} illustrates the scoping rules.

\paragraph{R1: Same-Scope Visibility}
The first semantic rule states that pods within the same scope can access each other's definitions. For example, the functions \texttt{a1} and \texttt{a2} can access each other because they are contained within the same scope A. This rule enables local collaboration between related code blocks within a module.

\paragraph{R2: Scope Encapsulation}
The second semantic rule is complementary to the first and establishes encapsulation boundaries. It states that pods are not visible outside their containing scope unless explicitly exported. For example, the function \texttt{c1} is not visible to the function \texttt{a2} because \texttt{c1} is contained within scope \texttt{C}, while \texttt{a2} is outside that scope. This encapsulation is vital for maintaining code integrity and preventing unintended interactions between functions in different scopes.

\paragraph{R3: Export Visibility}
Scopes are designed to be hierarchical, with higher-level functions implemented using lower-level functions. Therefore, we require a mechanism to expose public APIs to parent scopes. The third semantic rule introduces the concept of explicit exports: pods marked with the \texttt{\#export} tag become public APIs of their scope and are visible to the parent scope. The symbols of a scope consist of all exported symbols within that scope. For example, the function \texttt{c2} has an export tag, making it visible to \texttt{a2} in the parent scope A. Similarly, \texttt{a2} has an export tag, making it visible to \texttt{d1} in its parent scope D. Additionally, a scope itself can have an export tag, which exports all the scope's symbols as public APIs in the parent scope. For example, scope A has \texttt{\#export} tag and the symbol of A (\texttt{a2}) is exported as symbol of parent scope D. Now the symbol \texttt{a2} is visible outside scope D in function \texttt{e1}. This rule facilitates the integration of different modules, allowing developers to expose specific API functions from lower-level modules to higher-level modules.

\paragraph{R4: Lexical Scoping}
The fourth semantic rule implements lexical scoping, allowing functions to be visible to their descendant scopes. This feature is essential for writing utility functions or shared variables that are used across multiple modules. For example, the function \texttt{util1} is a utility function used in both Scope B and C, so we place it in the parent scope A, making it visible in both child scopes. The name and concept of lexical scoping are borrowed from traditional programming language semantics.

\paragraph{Composition of Rules}
The four semantic rules can be composed to handle more complex scenarios. For example, the function \texttt{c2} is visible in sibling scope pod \texttt{b2} because it is first exported to the parent scope with its export tag, and then becomes visible to child scope B via lexical scoping. Together, these scoping rules form the foundation of the hierarchical and modular code management system in CodePod, offering a flexible yet structured environment for software development.

\subsection{Formal Semantics of CodePod Scoping Rules}
\label{sec:formal-semantics}

To provide a rigorous foundation for CodePod's scoping system, we present a formal semantics that precisely defines the visibility and accessibility relationships between pods and scopes. This formalization enables precise reasoning about the behavior of CodePod's hierarchical scoping system and provides a basis for implementation correctness. Our semantics extends traditional lexical scoping concepts to support hierarchical modules with explicit export mechanisms, providing a mathematical framework for reasoning about program structure and dependencies.

\subsubsection{Syntax and Structure}

We begin by defining the syntactic structure of CodePod's hierarchical organization:

\begin{definition}[CodePod Structure]
A CodePod project $\mathcal{P}$ is a tuple $(S, P, \mathcal{H}, \mathcal{E}, \mathcal{X}, \mathcal{X}_S)$, where:
\begin{itemize}
    \item $S$ is a finite set of scopes
    \item $P$ is a finite set of pods
    \item $\mathcal{H} \subseteq S \times S$ is a hierarchical relation defining the parent-child relationship between scopes
    \item $\mathcal{E} \subseteq P \times S$ is the containment relation mapping pods to their containing scopes
    \item $\mathcal{X} \subseteq P \times P$ is the export relation defining which pods are exported from child scopes to parent scopes
    \item $\mathcal{X}_S \subseteq S \times S$ is the scope export relation defining which scopes are exported from child scopes to parent scopes
\end{itemize}
\end{definition}

\begin{definition}[Scope Hierarchy]
The scope hierarchy is a directed acyclic graph $(S, \mathcal{H})$ where:
\begin{itemize}
    \item $\mathcal{H}$ is irreflexive: $\forall s \in S, (s, s) \notin \mathcal{H}$
    \item $\mathcal{H}$ is transitive: if $(s_1, s_2) \in \mathcal{H}$ and $(s_2, s_3) \in \mathcal{H}$, then $(s_1, s_3) \in \mathcal{H}$
    \item $\mathcal{H}$ is antisymmetric: if $(s_1, s_2) \in \mathcal{H}$ and $(s_2, s_1) \in \mathcal{H}$, then $s_1 = s_2$
    \item There exists a unique root scope $s_{root} \in S$ such that $\forall s \in S \setminus \{s_{root}\}, \exists s' \in S: (s', s) \in \mathcal{H}$
\end{itemize}
\end{definition}

\begin{definition}[Well-formed CodePod Project]
A CodePod project $\mathcal{P} = (S, P, \mathcal{H}, \mathcal{E}, \mathcal{X}, \mathcal{X}_S)$ is well-formed if:
\begin{itemize}
    \item $(S, \mathcal{H})$ forms a valid scope hierarchy
    \item Every pod is contained in exactly one scope: $\forall p \in P, \exists! s \in S: (p, s) \in \mathcal{E}$
    \item Export relations respect scope hierarchy: if $(p_1, p_2) \in \mathcal{X}$, then $(\mathcal{E}(p_1), \mathcal{E}(p_2)) \in \mathcal{H}$
    \item Scope export relations respect scope hierarchy: if $(s_1, s_2) \in \mathcal{X}_S$, then $(s_1, s_2) \in \mathcal{H}$
\end{itemize}
\end{definition}

\subsubsection{Symbol Environment and Visibility}

We define the symbol environment and visibility relations that capture the essence of CodePod's scoping rules:

\begin{definition}[Symbol Environment]
A symbol environment $\Gamma$ is a partial function $\Gamma: S \times \Sigma \rightharpoonup P$, mapping scope-symbol pairs to the pod that defines the symbol in that scope, where $\Sigma$ is the set of all possible symbol names.
\end{definition}

\begin{definition}[Export Relation]
The export relation $\mathcal{X} \subseteq P \times P$ defines which pods are exported from child scopes to parent scopes:
\begin{itemize}
    \item $(p, p') \in \mathcal{X}$ if pod $p$ is exported from its scope to the scope containing pod $p'$
    \item This relation is derived from the \texttt{\#export} tags in the code
    \item The export relation respects scope hierarchy: if $(p_1, p_2) \in \mathcal{X}$, then $(\mathcal{E}(p_1), \mathcal{E}(p_2)) \in \mathcal{H}$
\end{itemize}
\end{definition}

\begin{definition}[Symbol Definition]
A symbol $\sigma \in \Sigma$ is defined in scope $s \in S$ by pod $p \in P$ if $\Gamma(s, \sigma) = p$ and $\mathcal{E}(p) = s$.
\end{definition}

\begin{definition}[Scope Export Relation]
The scope export relation $\mathcal{X}_S \subseteq S \times S$ defines which scopes are exported from child scopes to parent scopes:
\begin{itemize}
    \item $(s, s') \in \mathcal{X}_S$ if scope $s$ is exported to its parent scope $s'$
    \item This relation is derived from the \texttt{\#export} tag on the scope
    \item When a scope $s$ is exported to $s'$, all exported symbols of $s$ become public APIs in $s'$
\end{itemize}
\end{definition}

\subsubsection{Formal Scoping Rules}

We now formalize the four scoping rules (R1-R4) as mathematical relations:

\begin{definition}[Same-Scope Visibility (R1)]
For pods $p_1, p_2 \in P$:
\begin{align}
\text{visible}_{same}(p_1, p_2) \iff \mathcal{E}(p_1) = \mathcal{E}(p_2)
\end{align}
\end{definition}

\begin{definition}[Scope Encapsulation (R2)]
For pods $p_1, p_2 \in P$:
\begin{align}
\text{visible}_{encap}(p_1, p_2) \iff \mathcal{E}(p_1) = \mathcal{E}(p_2)
\end{align}
\end{definition}

\begin{definition}[Export Visibility (R3)]
For pods $p_1, p_2 \in P$:
\begin{align}
\text{visible}_{export}(p_1, p_2) \iff 
\left[
  (\mathcal{E}(p_1), \mathcal{E}(p_2)) \in \mathcal{H} \land (p_1, p_2) \in \mathcal{X}
\right] \\
\text{or}\quad
\left[
  (\mathcal{E}(p_1), \mathcal{E}(p_2)) \in \mathcal{X}_S
\right]
\end{align}
\end{definition}

\begin{definition}[Lexical Scoping (R4)]
For pods $p_1, p_2 \in P$:
\begin{align}
\text{visible}_{lexical}(p_1, p_2) \iff (\mathcal{E}(p_2), \mathcal{E}(p_1)) \in \mathcal{H}
\end{align}
\end{definition}

\subsubsection{Combined Visibility Relation}

The complete visibility relation combines all four rules:

\begin{definition}[Complete Visibility]
For any two pods $p_1, p_2 \in P$:
\begin{align}
\text{visible}(p_1, p_2) \iff \text{visible}_{same}(p_1, p_2) \lor \text{visible}_{export}(p_1, p_2) \lor \text{visible}_{lexical}(p_1, p_2)
\end{align}
\end{definition}

\begin{definition}[Visibility Properties]
The visibility relation satisfies the following properties:
\begin{itemize}
    \item \textbf{Reflexivity:} $\forall p \in P, \text{visible}(p, p)$ (every pod is visible to itself)
    \item \textbf{Transitivity:} If $\text{visible}(p_1, p_2)$ and $\text{visible}(p_2, p_3)$, then $\text{visible}(p_1, p_3)$
    \item \textbf{Antisymmetry:} If $\text{visible}(p_1, p_2)$ and $\text{visible}(p_2, p_1)$, then $p_1 = p_2$
\end{itemize}
\end{definition}

\subsubsection{Symbol Resolution Semantics}

We now define the formal semantics for symbol resolution in CodePod:

\begin{definition}[Symbol Resolution]
Given a symbol $\sigma \in \Sigma$ used in pod $p \in P$, the resolution function $\text{resolve}(\sigma, p)$ returns the defining pod $p' \in P$ such that:
\begin{align}
\text{resolve}(\sigma, p) = p' \iff \Gamma(\mathcal{E}(p'), \sigma) = p' \land \text{visible}(p', p)
\end{align}
If no such pod exists, $\text{resolve}(\sigma, p) = \bot$.
\end{definition}

\begin{definition}[Scope Chain]
For a scope $s \in S$, the scope chain $\text{chain}(s)$ is the sequence of scopes from $s$ to the root:
\begin{align}
\text{chain}(s) = [s_0, s_1, \ldots, s_n] \text{ where } s_0 = s, s_n = s_{root}, \text{ and } \forall i < n, (s_i, s_{i+1}) \in \mathcal{H}
\end{align}
\end{definition}

\subsubsection{Properties and Theorems}

We establish several important properties of the CodePod scoping system:

\begin{theorem}[Uniqueness of Symbol Resolution]
For any symbol $\sigma \in \Sigma$ and pod $p \in P$, if $\text{resolve}(\sigma, p)$ is defined, then it is unique.
\end{theorem}

\begin{proof}
By contradiction. Assume there exist two distinct pods $p_1, p_2 \in P$ such that $\text{resolve}(\sigma, p) = p_1$ and $\text{resolve}(\sigma, p) = p_2$. By definition, this means:
\begin{align}
\Gamma(\mathcal{E}(p_1), \sigma) = p_1 \land \text{visible}(p_1, p) \\
\Gamma(\mathcal{E}(p_2), \sigma) = p_2 \land \text{visible}(p_2, p)
\end{align}
Since $\Gamma$ is a function, if $\mathcal{E}(p_1) = \mathcal{E}(p_2)$, then $p_1 = p_2$, contradicting our assumption. If $\mathcal{E}(p_1) \neq \mathcal{E}(p_2)$, then by the scope hierarchy properties and visibility rules, this would create a cycle in the visibility relation, contradicting the acyclic nature of the scope hierarchy.
\end{proof}

\begin{theorem}[Transitive Visibility]
If $\text{visible}(p_1, p_2)$ and $\text{visible}(p_2, p_3)$, then $\text{visible}(p_1, p_3)$.
\end{theorem}

\begin{proof}
This follows from the transitivity of the scope hierarchy relation $\mathcal{H}$ and the definition of the visibility relations.
\end{proof}

\begin{theorem}[Export Propagation]
If pod $p$ is exported from scope $s_1$ to scope $s_2$, and $s_2$ is exported to scope $s_3$, then $p$ is visible in $s_3$.
\end{theorem}

\begin{proof}
By the definition of export visibility and the transitivity of the scope hierarchy, the export relation propagates through the hierarchy according to the export tags.
\end{proof}

\subsubsection{Operational Semantics}

We define the operational semantics for symbol resolution in terms of a lookup algorithm:

\begin{algorithm}[H]
\caption{Symbol Resolution Algorithm}
\begin{algorithmic}[1]
\Function{ResolveSymbol}{$\sigma$, $p$, $\Gamma$, $\mathcal{H}$, $\mathcal{E}$, $\mathcal{X}$}
    \State $s \gets \mathcal{E}(p)$ \Comment{Current scope}
    \State $chain \gets \text{chain}(s)$ \Comment{Scope chain to root}
    \For{$s_i \in chain$}
        \If{$\Gamma(s_i, \sigma)$ is defined}
            \State $p' \gets \Gamma(s_i, \sigma)$
            \If{$\text{visible}(p', p)$}
                \State \Return $p'$
            \EndIf
        \EndIf
    \EndFor
    \State \Return $\bot$ \Comment{Symbol not found}
\EndFunction
\end{algorithmic}
\end{algorithm}

\begin{theorem}[Correctness of Symbol Resolution]
The symbol resolution algorithm correctly implements the formal visibility relation:
\begin{align}
\text{ResolveSymbol}(\sigma, p, \Gamma, \mathcal{H}, \mathcal{E}, \mathcal{X}) = p' \iff \text{resolve}(\sigma, p) = p'
\end{align}
\end{theorem}

\begin{proof}
The algorithm searches through the scope chain in order, checking each scope for the symbol definition and verifying visibility according to the formal rules. The first valid definition found corresponds to the closest scope in the hierarchy that satisfies the visibility constraints.
\end{proof}

\subsubsection{Complexity Analysis}

\begin{theorem}[Symbol Resolution Complexity]
The time complexity of symbol resolution is $O(h \cdot |S|)$, where $h$ is the height of the scope hierarchy and $|S|$ is the number of scopes.
\end{theorem}

\begin{proof}
The algorithm traverses the scope chain, which has length at most $h$. For each scope, it performs a constant-time lookup in the symbol environment $\Gamma$. The total complexity is therefore $O(h \cdot |S|)$.
\end{proof}

\begin{theorem}[Incremental Update Complexity]
When a single pod is modified, the time complexity to update the symbol tables and resolve affected symbols is $O(k \cdot h)$, where $k$ is the number of symbols defined in the modified pod and $h$ is the height of the scope hierarchy.
\end{theorem}

\begin{proof}
Updating a pod requires: (1) removing old definitions from symbol tables ($O(k \cdot h)$ for propagation up the hierarchy), (2) adding new definitions ($O(k \cdot h)$ for propagation), and (3) resolving affected symbols in dependent pods ($O(k \cdot h)$ for each affected pod). Since the number of affected pods is typically small in practice, the overall complexity is $O(k \cdot h)$.
\end{proof}

This formal semantics provides a rigorous foundation for understanding and implementing CodePod's scoping system, ensuring that the intuitive rules R1-R4 are precisely defined and can be reasoned about mathematically. The complexity bounds establish that our approach scales efficiently with project size, making it suitable for large-scale development.


\subsection{Language-Agnostic Scoped Runtime via Source Level Transformation}
\label{sec:runtime}

\begin{figure}[ht]
    \centering
    \includegraphics[width=0.95\textwidth]{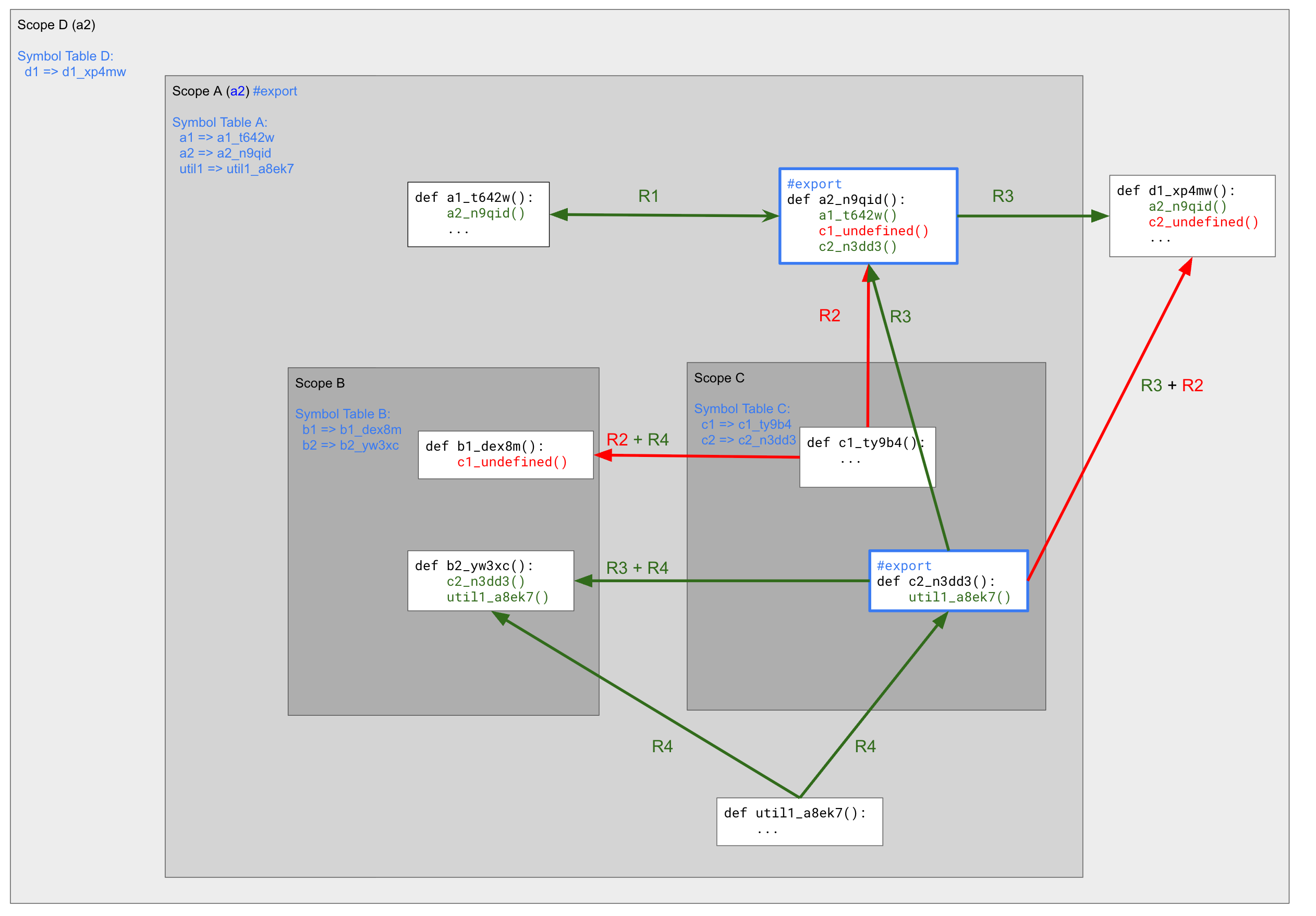}
    \caption{Source-level transformation pipeline: Code blocks are parsed to extract definitions and uses, symbol tables are constructed per scope, and code is rewritten with unique identifiers before kernel execution.}
    \label{fig:semantic-resolved}
\end{figure}

We now present the operational semantics of CodePod's runtime system, which implements the hierarchical scoping rules formalized in Section~\ref{sec:formal-semantics}. The key innovation of our runtime architecture is its language-agnostic design, which implements hierarchical scoping through systematic source-level transformations rather than kernel modifications.

Traditional module systems are tightly coupled to language-specific runtime environments, requiring modifications to interpreters, compilers, or virtual machines. This approach limits portability and requires significant implementation effort for each supported language. In contrast, CodePod's runtime operates as a transformation layer between user code and language kernels, implementing scoping semantics through systematic source code rewriting while preserving the interactive development experience.

The runtime system consists of three main phases: (1) a parsing and def-use analysis phase that extracts symbol definitions and references from abstract syntax trees; (2) a symbol table construction phase that builds hierarchical symbol environments according to our formal scoping rules; and (3) a code transformation phase that rewrites symbol references to implement the visibility relations defined in our semantics. This design ensures that our scoping process remains universal, applicable across different programming languages without requiring modifications to their respective interpreters or kernels.

Our language-agnostic design enables support for multiple programming languages through a unified transformation pipeline. We have implemented support for nine programming languages spanning different paradigms and features, as shown in Table~\ref{tab:programming-languages}. The implementation leverages tree-sitter for parsing and AST analysis, providing consistent symbol extraction across languages while maintaining language-specific syntax awareness.

The key insight underlying our approach is that hierarchical scoping can be implemented through AST-level transformations that are independent of language-specific runtime semantics. Each language requires only a configuration file defining the patterns for function definitions, class definitions, variable assignments, and function calls. This minimal language-specific configuration enables rapid addition of new language support while maintaining the theoretical guarantees of our formal semantics.

\subsubsection{\textbf{Runtime Architecture Intuitive Overview}}

The core insight of our runtime system is to implement hierarchical scoping through systematic source code transformation. Rather than modifying language kernels or implementing custom interpreters, we transform user code to respect scoping rules by rewriting symbol references with unique identifiers that encode both the symbol name and its defining context.

For each code block that defines a symbol, we append a unique pod identifier to create a globally unique name. For example, a function \texttt{a1} defined in pod with UUID \texttt{t642w} becomes \texttt{a1\_t642w}. This ensures that symbols with the same name in different pods do not conflict, while preserving the original symbol names for user comprehension.

For code blocks that reference symbols, we analyze the scoping rules to determine which definition is visible according to our formal semantics. If a symbol is visible, we rewrite the reference to use the unique identifier of the defining pod. If no visible definition exists, we rewrite the reference to include an "undefined" suffix, which will cause a runtime error when executed, providing immediate feedback about scoping violations.

The transformation process is illustrated in Figure~\ref{fig:semantic-resolved}. Consider a pod with UUID \texttt{n9qid} that defines a function \texttt{a2} calling three other functions: \texttt{a1}, \texttt{c1}, and \texttt{c2}. Before execution, the function name \texttt{a2} is rewritten as \texttt{a2\_n9qid}. According to our scoping rules, \texttt{a1} and \texttt{c2} are visible (via rules R1 and R3 respectively), while \texttt{c1} is not visible. Therefore, the callsites are resolved to \texttt{a1\_t642w()} and \texttt{c2\_n3dd3()}, while \texttt{c1} is rewritten to \texttt{c1\_undefined()} and will throw an error at runtime. The rewritten code blocks are then sent to the language kernel for execution, ensuring that the runtime behavior adheres to our formal scoping semantics.

The runtime system operates through three sequential phases:

\begin{enumerate}
    \item \textbf{Def-use Analysis:} We parse code blocks to extract (1) definitions of top-level symbols (functions, variables, classes) and (2) uses of undefined variables. For example, pod a1 has \texttt{\{def: a1, use: a2\}}, while pod a2 has \texttt{\{def: a2, use: a1, c1, c2\}}.
    \item \textbf{Symbol Table Construction:} We build a symbol table for each scope to record the symbols defined in that scope. For example, the symbol table for scope A contains \texttt{\{a1: a1\_t642w, a2: a2\_n9qid, util1: util1\_a8ek7\}}, while the symbol table for scope C contains \texttt{\{c1: c1\_ty9b4, c2: c2\_n3dd3\}}. The symbol table is maintained incrementally during the user edit session.
    \item \textbf{Symbol Resolution and Rewriting:} Before a code block is sent to language kernels for execution, we resolve symbol uses in the symbol table and rewrite them with the resolved symbol names including the target pod UUID. For example, the callsite of a1 in a2 is rewritten to \texttt{a1\_t642w}, and unresolved references are rewritten to \texttt{c1\_undefined()}. The rewritten code blocks are then sent to the language kernel for execution.
\end{enumerate}

We now present the detailed algorithms for each phase.

\subsubsection{\textbf{Phase 1: Parsing and Def-use Analysis}}

We employ the tree-sitter parsing library for parsing source code and extracting symbol definitions and references. Tree-sitter provides high-performance incremental parsing with linear time complexity for initial parsing and sublinear complexity for incremental updates. The parsing step runs once at initialization and is re-executed only when code changes occur.

\begin{algorithm}[ht]
    \small
    \caption{Pattern-Matching Based Def-use Analysis (Phase 1) \label{alg:parse}}
    \begin{algorithmic}[1]
    \small
    \State t1 $\gets$ \texttt{(module (func\_def (identifier) \textcolor{red}{@func}))}
    \State t2 $\gets$ \texttt{(module (class\_def (identifier) \textcolor{teal}{@class}))}
    \State t3 $\gets$ \texttt{(call (identifier) \textcolor{cyan}{@callsite})}
    \State t4 $\gets$ \texttt{(module (assign\_stmt (identifier) \textcolor{blue}{@vardef}))}
    \State t5 $\gets$ \texttt{((identifier) \textcolor{orange}{@varuse})}
    \Function{ParsePod}{pod}
        \State parser $\gets$ TreeSitter(lang)
        \State syntaxTree $\gets$ parser.parse(pod.code) \Comment{Parse using Tree-sitter}
        \Statex \Comment{Pattern matching for defs and uses}
        \State functions $\gets$ match(syntaxTree, t1) \Comment{Top level function defs}
        \State classes $\gets$ match(syntaxTree, t2) \Comment{Top level class defs}
        \State callsites $\gets$ match(syntaxTree, t3) \Comment{function calls}
        \State vardefs $\gets$ match(syntaxTree, t4) \Comment{Top level var defs}
        \State varuses $\gets$ match(syntaxTree, t5) \Comment{variable/symbol uses}
        \State varuses $\gets$ \Call{DefUse}{syntaxTree} \Comment{Alternative precise def-use alg} 
        \State defs $\gets$ functions $\cup$ classes $\cup$ vardefs
        \State uses $\gets$ callsites $\cup$ varuses $\setminus$ defs
        \State \Return syntaxTree, defs, uses
    \EndFunction

    \Statex
    \State cache $\gets$ \texttt{empty dict}
    \Statex

    \Function{IncrementalParse}{pod, changes}
    \If {cache.get(pod.UUID) is None}
        \State syntaxTree, defs, uses $\gets$ \Call{ParsePod}{Pod}
        \State cache.set(pod.UUID, syntaxTree, defs, uses)
    \Else
        \State syntaxTree, defs, uses = cache.get(pod.UUID)
    \EndIf
    \State updatedSyntaxTree $\gets$ Apply(changes, syntaxTree) \Comment{incremental}
    \State \_, defs, uses $\gets$ \Call{ParsePod}{Pod}
    \State \Return updatedSyntaxTree, defs, uses
    \EndFunction
    \end{algorithmic}
\end{algorithm}

We implement def-use analysis through two complementary approaches. The primary approach uses pattern-matching with tree-sitter queries, as shown in Algorithm~\ref{alg:parse}. This method employs simple tree-sitter query strings similar to regular expressions to identify function definitions, class definitions, function calls, variable definitions, and variable uses. These patterns are largely language-agnostic, requiring only minimal adjustments for different programming languages.

For scenarios requiring more precise def-use analysis, we provide an alternative recursive descent implementation shown in Algorithm~\ref{alg:detailed-parse}. This approach traverses the AST systematically, tracking defined symbols and identifying unbound references. While more complex and language-specific, this method provides more accurate def-use relationships for languages with complex scoping rules.

\begin{algorithm}[ht]
    \caption{Alternative Precise Def-use Analysis (Phase 1) \label{alg:detailed-parse}}
    \begin{algorithmic}[1]
    \small
    \Function{DefUse}{syntaxTree}
        \State defined $\gets \emptyset$ \Comment{Set of currently defined symbols}
        \State unbound $\gets \emptyset$ \Comment{Set of symbols used before being defined}
        \State \Call{AnalyzeBlock}{syntaxTree, defined, unbound}
        \State \Return unbound
    \EndFunction
    \Function{AnalyzeBlock}{block, defined, unbound}
        \ForAll{node in block}
            \If{node is AssignStmt}
                \State \Call{AnalyzeAssign}{node, defined, unbound}
            \ElsIf{node is Function}
                \State \Call{AnalyzeFunction}{node, defined, unbound}
            \ElsIf{node is If}
                \State \Call{AnalyzeIf}{node, defined, unbound}
            \ElsIf{node is For or While}
                \State \Call{AnalyzeLoop}{node, defined, unbound}
            \EndIf
            \State $\cdots$
        \EndFor
    \EndFunction
    \Function{AnalyzeAssignStmt}{node, defined, unbound}
        \State \Call{AnalyzeExpr}{node.rhs, defined, unbound}
        \State defined $\gets$ defined $\cup$ node.lhs
    \EndFunction
    \Function{AnalyzeFunction}{node, defined, unbound}
        \State localDefined $\gets$ defined $\cup \{node.name\}$
        \State \Call{AnalyzeBlock}{node.body, localDefined, unbound}
    \EndFunction
    \Function{AnalyzeIf}{node, defined, unbound}
        \State \Call{AnalyzeExpr}{node.condition, defined, unbound}
        \State \Call{AnalyzeBlock}{node.body, defined, unbound}
        \State \Call{AnalyzeBlock}{node.else, defined, unbound}
    \EndFunction
    \State $\cdots$
    \end{algorithmic}
\end{algorithm}

\subsubsection{\textbf{Phase 2: Symbol Table Construction}}

Following def-use analysis, we construct and maintain hierarchical symbol tables that implement our formal scoping semantics. Each scope maintains a symbol table recording the symbols defined within that scope, including public symbols exported from child scopes according to rule R3.

\begin{algorithm}[t]
\caption{Symbol Table Construction (Phase 2) \label{alg:symbol-table}}
\begin{algorithmic}[1]
\small
\small

\Function{BuildSymbolTables}{pod}
    \State defs $\gets$ \Call{ParsePod}{pod}
    \State oldDefs $\gets$ pod.oldDefs
    \State symbolTable $\gets$ pod.scope.symbolTable
    \ForAll{def in oldDefs}
        \State symbolTable.remove(def)
    \EndFor
    \ForAll{def in defs}
        \State symbolTable.add(def)
    \EndFor
    \State pod.oldDefs $\gets$ defs
    \If{pod.isPublic}
        \State \Call{PropagateToParent}{pod, defs}
    \EndIf
\EndFunction

\Statex
\Function{PropagateToParent}{pod, defs}
    \State parentScope $\gets$ pod.scope.parent
    \If{parentScope is not ROOT}
        \State parentScope.symbolTable.add(defs)
        \State \Call{PropagateToParent}{parentScope, defs}
    \EndIf
\EndFunction

\end{algorithmic}
\end{algorithm}

The symbol table construction algorithm, shown in Algorithm~\ref{alg:symbol-table}, operates incrementally. For each pod, we first retrieve the symbol definitions from the parsing phase. We then update the scope's symbol table by removing old definitions and adding new ones, implementing rule R1. If the pod is marked as public, we recursively propagate the definitions to parent scopes, implementing rule R3. This propagation continues until reaching the root scope.

The time complexity for building symbol tables is $O(n k \log(n))$ where $n$ is the number of pods and $k$ is the average number of definitions per pod. The $\log(n)$ factor accounts for propagation up the scope hierarchy. Once symbol tables are constructed, maintaining them for a single pod modification requires only $O(k \log(n))$ time.

\subsubsection{\textbf{Phase 3: Symbol Resolution and Code Rewriting}}

The final phase implements symbol resolution according to our formal semantics and rewrites code to ensure runtime behavior matches our scoping rules. Each pod receives a unique identifier that is appended to symbol names to create globally unique identifiers.

\begin{algorithm}[t]
\caption{Symbol Resolution and Code Rewriting (Phase 3) \label{alg:resolve}}
\begin{algorithmic}[1]
\small

\Function{ResolveSymbol}{symbolUse, pod}
    \State localSymbolTable $\gets$ pod.scope.symbolTable
    \State varuses $\gets$ \Call{ParsePod}{pod}
    \State unresolved $\gets$ varuses
    \State unresolved $\gets$ unresolved $\setminus$ localSymbolTable \Comment{Resolving}
    \State resolved $\gets$ \Call{ResolveUp}{pod, unresolved}
    \State \Return resolved, unresolved
\EndFunction

\Statex
\Function{ResolveUp}{node, unresolved}
    \State resolved $\gets$ \texttt{empty list}
    \State scope $\gets$ node.scope
    \State unresolved $\gets$ unresolved $\setminus$ scope.symbolTable \Comment{Resolving}
    \If{unresolved is empty or scope is ROOT}
        \State \Return resolved
    \EndIf
    \State \Return \Call{ResolveUp}{scope.parent, unresolved}
\EndFunction

\Statex

\Function{RewriteCode}{pod}
    \State defs, uses $\gets$ \Call{ParsePod}{pod}
    \State resolved, unresolved $\gets$ \Call{ResolveSymbol}{pod}
    \State rewrittenCode $\gets$ \texttt{pod.code}
    \ForAll{name, loc in defs}
    \Statex \Comment{Postfix UUID for symbol defs}
        \State rewrittenCode.replace(loc, name + pod.UUID)
    \EndFor
    \ForAll{name, loc, originPod in resolved} 
    \Statex \Comment{Replace symbol uses with resolved target UUID}
        \State rewrittenCode.replace(loc, name + originPod.UUID)
    \EndFor
    \ForAll{name, loc in unresolved}
        \State rewrittenCode.replace(loc, name + "undefined")
        \State \texttt{raise UnresolvedSymbolError}
    \EndFor
    \State \Return rewrittenCode
\EndFunction

\end{algorithmic}
\end{algorithm}

The symbol resolution algorithm, shown in Algorithm~\ref{alg:resolve}, implements our formal visibility relations. For each symbol use, we first check the local scope's symbol table (implementing rules R1, R2, and R3). If the symbol cannot be resolved locally, we recursively search up the scope hierarchy (implementing rule R4). This recursive algorithm provides lexical scoping with proper symbol shadowing support.

The code rewriting phase replaces symbol definitions with unique identifiers and resolves symbol uses according to the visibility analysis. Unresolved symbols are rewritten to include an "undefined" suffix, which will cause runtime errors and provide immediate feedback about scoping violations.

The time complexity for rewriting a single code block is $O(k \log(n))$ where $k$ is the average number of symbol uses per pod and $\log(n)$ accounts for hierarchical symbol resolution.

This three-phase runtime system effectively implements our formal scoping semantics in a language-agnostic manner, enabling hierarchical organization across multiple programming languages while maintaining the interactive development experience that makes notebooks valuable. The system has been implemented for Python, JavaScript, Julia, and Racket, demonstrating its applicability across diverse language paradigms.


\subsection{Def-use Visualization}
\label{sec:defuse}

Understanding code dependencies is essential for effective software development. CodePod automatically analyzes and visualizes definition-use relationships between code blocks, providing immediate insight into program structure and data flow. This visualization is built during the symbol resolution phase described in Section~\ref{sec:canvas}, where we identify which symbols are defined in each pod and which pods reference those symbols.

The def-use visualization renders directed edges on the canvas to depict these relationships, helping users understand dependencies and arrange code blocks logically. For example, users can position definition pods in the top-left and arrange usage pods toward the bottom-right, creating a natural flow that reflects the program's logical structure.

The visualization respects scope boundaries: edges are only shown within a single scope to avoid visual clutter. When a code block references a symbol defined in another scope, the edge connects from the referencing block's ancestor scope to the defining block's ancestor scope at their least common ancestor in the scope hierarchy. This approach maintains visual clarity while accurately representing cross-scope dependencies.

\subsection{Execution Order Management}

Traditional Jupyter notebooks enforce a linear execution model where code cells are executed sequentially from top to bottom. CodePod's hierarchical 2D organization enables more sophisticated execution strategies that can be combined flexibly based on user needs and project requirements.

\begin{enumerate}
    \item \textbf{Individual Execution:} Like Jupyter, users can execute any individual code block independently. This capability is crucial for interactive and exploratory programming, allowing developers to test specific functions or examine intermediate results without running the entire project.
    
    \item \textbf{Dependency-Based Ordering:} CodePod automatically computes topological ordering based on the def-use relationships identified during symbol resolution. When executing multiple pods, the system ensures that dependencies are satisfied by executing definition pods before their usage pods. For cyclic dependencies (e.g., mutually recursive functions), the system detects cycles and executes one dependency arbitrarily to break the cycle.
    
    \item \textbf{Manual Dependency Specification:} Def-use analysis may not capture all execution dependencies, particularly for procedural scripts where code blocks perform sequential operations without explicit variable dependencies. Users can manually add dependency edges between pods and scopes to specify execution ordering requirements. This feature enables CodePod to handle complex workflows that extend beyond simple function call relationships.
    
    \item \textbf{Spatial Ordering:} When no explicit dependencies are specified, CodePod defaults to geographical ordering based on the (x,y) coordinates of pods on the canvas, executing from top-left to bottom-right. This provides predictable behavior while leveraging the spatial organization of the 2D interface.
\end{enumerate}

Users can execute code through multiple modalities: individual pod evaluation for interactive development, chain execution following dependency edges for systematic testing, or scope execution that processes all descendant code blocks and scopes according to the defined execution order. This flexibility makes CodePod a superset of Jupyter's execution model, supporting both the familiar interactive development experience and more sophisticated project management workflows.


\section{Implementation and Evaluation}
\label{sec:eval}

CodePod is built on a nested 2D canvas architecture that supports zooming and panning for both detailed code examination and high-level project overview. The interface includes several navigation and organization features: (1) symbol tables displayed beside the canvas and each scope showing private functions and public APIs, enabling users to quickly identify symbol definitions and jump to their locations; (2) defined symbol names prominently displayed atop each pod in larger font for enhanced visibility; (3) a table of contents for navigating scopes and pods with jump-to-pod functionality; and (4) a comprehensive search feature for locating symbols across the project.

For large-scale projects, CodePod supports subpages—additional canvases for submodules that maintain the same semantic behavior as scopes while avoiding performance overhead from displaying too many pods on a single page. The system also enables repository-level modularity, allowing CodePod repositories to serve as libraries that can import functions and classes into other repositories, further enhancing modularity and scalability. Version control integration is provided through direct git command passthrough, maintaining compatibility with existing development workflows.

Figure~\ref{fig:tool-interface} shows a screenshot of CodePod with a real-world project implementing a PyTorch neural network classifier for the Fashion-MNIST dataset—the same project depicted in Figure~\ref{fig:mnist-jupyter}. This example demonstrates typical CodePod usage, where different scopes group related components, and def-use visualization edges show dependencies both within and between scopes. The code blocks are organized into three distinct modules: data loading, model training, and evaluation. Users can clearly visualize function relationships and the complete training pipeline. Code pods execute with results displayed below them, maintaining Jupyter's interactive development experience while providing faster development cycles through hierarchical organization.

\begin{table}[ht]
    \footnotesize
    \centering
    \begin{tabular}{l|c|l}
    \toprule
    Language & Kernel & Features \\ \midrule
    Python & ipython~\cite{ipython_kernel} & data science, ML/AI, web dev, automation \\ 
    JS/TS & deno~\cite{deno_kernel} & web frontend/backend, Node.js ecosystem, async/await \\
    Julia & IJulia~\cite{ijulia_kernel} & scientific computing, multiple dispatch, macros \\ 
    Racket  & iracket~\cite{iracket_kernel} & functional programming, hygienic macros, DSLs \\ 
    Ruby & iruby~\cite{iruby_kernel} & web development (Rails), metaprogramming, elegant syntax \\ 
    C\# & dotnet-interactive~\cite{dotnet_interactive_kernel} & .NET ecosystem, Windows development, game dev (Unity) \\ 
    C++  & xeus-cling~\cite{xeus_cling_kernel} & systems programming, performance-critical apps, templates \\ 
    Go & gophernotes~\cite{gophernotes_kernel} & cloud services, microservices, goroutines, simplicity \\ 
    Rust & evcxr\_jupyter~\cite{evcxr_jupyter_kernel} & memory safety, systems programming, zero-cost abstractions \\ \bottomrule
    \end{tabular}
    \caption{CodePod's Language-Agnostic Runtime Supports Diverse Programming Languages}
    \label{tab:programming-languages}
\end{table}

We have implemented support for nine major programming languages representing diverse paradigms including compiled and interpreted languages, static and dynamic type checking, and low-level and high-level programming approaches. This breadth demonstrates the language-agnostic nature of our runtime system.

Our implementation is available as open-source software for research purposes and reproducibility.
The tool, source code, user study data, and evaluation benchmarks are available at \url{https://codepod.io}.

\subsection{User Study Design and Methodology}

We conducted a controlled qualitative user study to evaluate CodePod's effectiveness compared to Jupyter in real-world software development tasks. Our study employed a mixed-methods approach combining qualitative feedback with quantitative metrics to provide comprehensive insights into developer experience and tool effectiveness.

\subsubsection{Participants}

We recruited seven senior software engineers and research scientists from technology companies, with an average of 10 years of programming experience and regular Jupyter usage (multiple times per week). We focused on senior developers because they have extensive experience with development tools, are more likely to encounter the scalability challenges that CodePod addresses, and can provide valuable feedback on professional adoption potential.

\begin{table}[t]
    \footnotesize
    \centering
    \begin{tabular}{r|l|l|l}
    \toprule
        Developer & Education & Current Role & Jupyter Experience \\
        \midrule
        1 & EE PhD (5+ years) & FAANG ML Scientist (3.5 yrs) & Multiple times/week \\
        2 & CS PhD (5+ years) & FAANG RL Scientist (1.5 yrs) & Multiple times/week \\
        3 & CS MS (5+ years) & FAANG ML Engineer (5 yrs) & Multiple times/week \\
        4 & CS PhD (5+ years) & NLP Scientist (1.5 yrs) & Multiple times/week \\
        5 & CS MS (2+ years) & FAANG Research Scientist (6 yrs) & Multiple times/week \\
        6 & CS PhD Candidate (5+ years) & NLP/LLM Researcher (6th yr) & Multiple times/week \\
        7 & CS MS (2+ years) & NLP Scientist (4 yrs) & Multiple times/week \\
        \bottomrule
    \end{tabular}
    \caption{Participant demographics and background information}
    \label{tab:participant-background}
\end{table}

As shown in Table~\ref{tab:participant-background}, participants represent diverse experience levels from graduate students to senior researchers and engineers. All work in machine learning, natural language processing, or related fields—aligning with Jupyter's primary user base. Their backgrounds span both academic and industrial settings, with several at major technology companies (FAANG) and others at research institutions. This diversity ensures our evaluation captures perspectives from different development contexts and use cases.

\subsubsection{Study Design}

We began with a 30-minute training session introducing CodePod's major features and demonstrating the tool. Participants then worked with three real-world Python projects in both Jupyter and CodePod environments.

Rather than asking participants to develop projects from scratch, we provided existing code for them to read, run, and understand. This approach was chosen for three reasons: (1) full project development would require many hours, making participant recruitment difficult; (2) each participant could only develop a task once, making benchmarks ineffective if sessions didn't run smoothly; and (3) our goal was to help participants understand CodePod's features well enough to provide meaningful comparisons with Jupyter.

We collected both qualitative feedback through open-ended questions and quantitative metrics through structured questionnaires to triangulate our findings.

\subsubsection{Benchmark Projects}

We selected three real-world Python projects spanning different domains and complexity levels:

\begin{table}[t]
    \small
    \centering
    \begin{tabular}{l|r|r}
    \toprule
        Project & \# Code Blocks & Lines of Code (LoC) \\
        \midrule
        PyTorch Fashion-MNIST & 15 & 130\\
        Review Sentiment Analysis & 40 & 133\\
        Python Web Scraping & 20 & 110\\
        \bottomrule
    \end{tabular}
    \caption{Statistics of Three Benchmark Projects}
    \label{tab:benchmarks}
\end{table}

The first project adapts the well-known PyTorch tutorial for training a neural network classifier on the Fashion-MNIST dataset. The second project uses NLTK and pretrained language models for sentiment analysis on Amazon food reviews, adapted from a highly-rated Kaggle competition notebook. The third project is a Python web scraper for quotes and books. These benchmarks are available in the supplementary materials.

After completing all three projects on CodePod, we confirmed that all participants had sufficient understanding of the tool to provide meaningful evaluations and comparisons with Jupyter.

\subsection{Q1: Perceived Scalability and Project Management}

To understand how developers perceive the scalability of both tools, we asked participants to evaluate how many code blocks they could comfortably develop in Jupyter versus CodePod, along with their reasoning for these assessments.



\begin{table}[t]
    \small
    \centering
    \begin{tabular}{r|r|r|r}
    \toprule
        User & \#blocks in Jupyter & \#blocks in CodePod & Improvement Factor \\
        \midrule
        1 & 20 & 80 & 4.0x \\
        2 & 30 & 50-80 & 1.67-2.67x \\
        3 & 15 & 20 &  1.33x\\
        4 & 15 & 15 & 1.0x\\
        5 & 20 & 25 & 1.25x\\
        6 & 50 & 130 & 2.6x\\
        7 & 100 & 20 & 0.2x\\
        \bottomrule
    \end{tabular}
    \caption{Q1: Number of code blocks users can comfortably develop in Jupyter and CodePod}
    \label{tab:q1}
\end{table}

The qualitative feedback reveals diverse perspectives on scalability, with participants providing rich insights into their reasoning. As shown in Table~\ref{tab:q1}, most developers (5 out of 7) perceive CodePod as offering significant scalability improvements over Jupyter, with estimates ranging from 1.33x to 4x improvement. Their qualitative feedback highlights several key factors contributing to this perceived scalability.

Participants who found CodePod more scalable emphasized the benefits of seeing the overall code structure and the organizational clarity provided by nested scopes. For instance, one participant noted that "the 2D canvas can visualize the structure of the code" and that "for long-term code maintenance, CodePod is much easier for users to view the overall structure of code and recall the function of each part of code." Another participant highlighted how "nested scopes help them understand the code by pieces."

However, the qualitative data also reveals important counter-perspectives. Two developers expressed reservations about CodePod's scalability benefits. One participant (Developer \#4) perceived no difference between the tools, while another (Developer \#7) actually preferred Jupyter's scalability. Their feedback provides valuable insights into potential limitations and user preferences.

The qualitative responses reveal that some developers use Jupyter primarily for drafting and exploratory work, where the linear structure serves their needs well. As one participant noted, "I use Jupyter for drafting, which is meant to be messy, and the canvas doesn't make things easier." Another participant expressed a general preference against canvas-based tools, comparing coding on a canvas to note-taking on canvas and stating that they "don't like graphical note-taking apps."

These diverse perspectives suggest that CodePod's value proposition may vary significantly based on individual work styles and project types. While most participants found the hierarchical organization beneficial for larger, more structured projects, some preferred Jupyter's simplicity for quick prototyping and exploration.

\subsection{Q2: Feature Ablation Study}

\begin{table}[t]
    \small
    \centering
    \begin{tabular}{r|r|r|r|r|r}
    \toprule
        Developer &  \makecell{Jupyter \\ (baseline)}  & Overall & 2D Canvas & \makecell{Hierarchical\\ Scopes} & \makecell{Def-use\\ visualization} \\
        \midrule
        1 & 5 & 8.5 & 8.5 & 9 & 8.5\\
        2 & 5 & 8 & 7 & 9 & 10\\
        3 & 5 & 8 & 8 & 7 & 8\\
        4 & 5 & 7 & 6 & 8 & 6 \\
        5 & 5 & 6 & 3 & 7 & 6 \\
        6 & 5 & 9 & 9 & 10 & 8 \\
        7 & 5 & 3 & 2 & 5 & 5 \\
        \midrule
        avg & 5 & 7.071 & 6.214 & 7.857 & 7.357 \\
        \bottomrule
    \end{tabular}
    \caption{Q2: Ablation Study: rating (0-10) of different features.}
    \label{tab:q2}
\end{table}

We conducted an ablation study to understand the relative usefulness of CodePod's three main features: 2D canvas, hierarchical scopes and semantics, and def-use visualization. Participants rated each feature on a 0-10 scale, with 5 representing Jupyter's baseline performance. A score of 0 indicates CodePod is significantly worse than Jupyter, while 10 indicates CodePod is significantly better.

Specifically, we asked users to rate the following aspects:

\begin{enumerate}
    \item The overall usefulness of CodePod.
    \item The usefulness of the 2D canvas interface and navigation features, including the ability to freely arrange code blocks (e.g., side-by-side rather than linearly) and to zoom in or out for both detailed and high-level views.
    \item The usefulness of nested scopes and semantic scoping rules, such as grouping related code blocks for clearer organization, supporting hierarchical nesting, and providing namespace separation and API exports during code execution.
    \item The usefulness of def-use (definition-use) visualization.
\end{enumerate}

Results are shown in Table~\ref{tab:q2}. Most users (6 of 7) rate CodePod overall better than Jupyter, with 4 of 7 giving very high scores of 8 or higher. All three features are shown to be useful, with hierarchical scopes being the most popular feature, averaging a score of 7.9.

Some users expressed negative opinions about the 2D canvas interface, but based on their feedback, this appears to be more of a personal preference. For example, developer 5 stated they are "terrible at this 2D canvas thing!"

\subsection{Q3: Migration Effort Assessment}

\begin{table}[t]
    \small
    \centering
    \begin{tabular}{r|r|r|r}
    \toprule
        Developer &  Fashion-MNIST & Amazon Food Review & Web Scraping \\
        \midrule
        1 & 3  & 8  & 4  \\
        2 & 5 & 10 & 5 \\
        3 & 10 & 20 & 15  \\
        4 & 10 & 20 & 15 \\
        5 & 10 & 15 & 10 \\
        6 & 15 & 15 & 12 \\
        7 & 10 & 15 & 15 \\

        \bottomrule
    \end{tabular}
    \caption{Q3: Time spent (minutes) to organize the pods on CodePod}
    \label{tab:q3}
\end{table}

We asked participants to estimate the time required to migrate and organize Jupyter notebooks to CodePod to assess whether this introduces significant overhead. The migration process involves importing notebooks linearly, then using def-use edges to guide layout on the 2D canvas. When necessary, developers group related code into modules, embed lower-level modules inside higher-level ones, and mark interfaces with \texttt{\#export} tags.

Results are shown in Table~\ref{tab:q3}. Overall, developers are optimistic about the migration effort, which can typically be completed within 10 to 15 minutes. Participants shared insights about factors affecting migration time, noting that well-structured code with clear functions requires less effort. For example, developer 1 reported being "very familiar with the MNIST classifier domain and thus it's easier to work with the code." However, the same developer found the sentiment analysis project more challenging because the code was "not organized into functions, scattered code snippets across variables and scripts."

\subsection{Q4: Navigation Efficiency}

\begin{table*}[t]
    \small
    \centering
    \begin{tabular}{r|>{\raggedleft\arraybackslash}p{1.5cm}|>{\raggedleft\arraybackslash}p{1.5cm}|>{\raggedleft\arraybackslash}p{1cm}|>{\raggedleft\arraybackslash}p{2cm}|>{\raggedleft\arraybackslash}p{2cm}|>{\raggedleft\arraybackslash}p{1cm}}
    \toprule
    Developer & \multicolumn{3}{c|}{Distance of co-edit blocks (jumps)} & \multicolumn{3}{c}{Time to navigate to block-to-edit (seconds)} \\
    \midrule
    & Jupyter         & CodePod         & Factor       & Jupyter        & CodePod        & Factor \\
    \midrule
    1 &        7   & 2          & 3.5x         &    20       &    5       & 4.0x \\
    2 &        5   &  4         & 1.25x        &    3       &   1       & 3.0x \\
    3 & 4  & 1 & 4.0x & 5 & 1 & 5.0x \\
    4 & 4 & 2 & 2.0x & 3 & 1 & 3.0x \\
    5 & 4 & 2 & 2.0x & 10 & 10 & 1.0x \\
    6 & 10 & 1 & 10.0x & 60 & 1 & 60.0x \\
    7 & - & - & - & - & - & - \\
    \bottomrule
    \end{tabular}
    \caption{Q4: Time and distance to navigate code blocks. Developer \#7 didn't feel comfortable providing feedback for this question.}
    \label{tab:q4}
\end{table*}

Navigation efficiency is a critical factor in development productivity, particularly for Jupyter notebooks where code blocks are linearly ordered. One participant noted that "using Jupyter, I need to update code that's far away. I spend a lot of time navigating among pods, editing, running, then navigating back to the current code. I spend lots of time jumping around."

We evaluated whether CodePod improves navigation efficiency using two metrics: the distance between related code blocks and the time required to find and navigate to target blocks for editing.

As shown in Table~\ref{tab:q4}, most users report that CodePod significantly reduces navigation effort, with improvements ranging from 2x to 10x in distance reduction and 1x to 60x in time savings. These improvements are attributed to the canvas model and hierarchical organization with visualized dependencies. Developer \#6 reported the most dramatic improvement—a 60x reduction in navigation time—because "the code structure becomes very clear after careful organization on CodePod."

\subsection{Threats to Validity}

We acknowledge several limitations in our evaluation that may affect the generalizability of our findings:

\textbf{Internal Validity:} Our user study involved seven participants, which limits the statistical significance of our results. However, the small cohort provided diverse perspectives from senior developers with varying preferences and use cases. The qualitative nature of our study provides rich insights into developer experience, but future work should include larger-scale studies to strengthen statistical validity.

\textbf{External Validity:} Participants were recruited through personal networks and referrals, which may introduce selection bias. To mitigate this, we explicitly instructed all participants to evaluate the tool objectively and provided standardized evaluation criteria. Additionally, two of the three benchmark projects involve machine learning, which reflects Jupyter's primary user base but may limit generalizability to other domains.

\textbf{Construct Validity:} Our evaluation emphasized efficiency-related metrics (scalability, navigation time, organization effort) which may not fully capture the richness of the developer experience. To address this, we complemented quantitative findings with qualitative feedback and open-ended responses that provide deeper insights into user preferences and pain points.

\textbf{Learning Effects:} Participants had limited time to familiarize themselves with CodePod, which may have affected their performance and evaluation. While we provided training sessions, the learning curve for the 2D canvas interface could have influenced results. Future studies should include longer-term usage scenarios to account for learning effects.

Despite these limitations, our multi-faceted evaluation approach provides strong evidence for the effectiveness of our approach, combining theoretical analysis with empirical validation and large-scale testing.

\subsection{Large-Scale Validation}

To validate CodePod's effectiveness beyond the scope of typical notebooks, we conducted a large-scale evaluation using real-world software projects. This evaluation demonstrates that our hierarchical scoping model naturally extends to large-scale development scenarios.

We chose two popular real-world Python projects: the Scrapy project, a web scraping framework, and the FastAPI project, a web framework for building APIs with Python. Both projects are well-known and widely used in the Python community, with 54.5k and 82k GitHub stars respectively. Scrapy has 164 Python files and 17,915 lines of code, while FastAPI has 42 Python files and 9,921 lines of code.

We converted the two projects into CodePod as follows. First, we built a file dependency graph by analyzing the importing relations between files. For each file, we split the file into code pods, each containing a function or a class. We put each file into a scope and built the hierarchy of the scopes using the file dependency graph. If a function is imported into other files, we marked it public with the \texttt{\#export} tag.

After the conversion, we completed two validation tasks. First, we migrated all unit tests onto CodePod, placing tests alongside the functions they test, and verified that all tests passed successfully. This demonstrates that our transformation preserves program semantics and that the hierarchical organization does not affect program correctness.

Second, we validated CodePod's effectiveness for real-world development tasks by implementing GitHub pull requests (PRs). PRs represent typical software development activities involving changes across multiple functions and files. We selected 5 PRs for each project, checked out the code before each PR, verified the presence of bugs, implemented the PR changes in CodePod, and confirmed that the bugs were fixed. We successfully implemented all 10 PRs, demonstrating that CodePod is suitable for real-world software development activities.

This large-scale validation provides strong evidence that our hierarchical scoping model extends naturally beyond notebook boundaries to support large-scale software development, bridging the gap between exploratory development and production software engineering.

\section{Related Work}
\label{sec:related}

Our work builds upon and extends research in interactive development environments, spatial code organization, and module systems. We position CodePod within the broader context of programming language research and software engineering tools.

\subsection{Extensions to Notebook Environments}
Our work is inspired by the Jupyter Notebook~\cite{jupyter} and introduces scalability extensions. Jupyter Notebook is a web-based interactive notebook. Jupyter notebooks consist of code cells that can be run in arbitrary order, with results displayed underneath. This supports literate programming by combining code, text, and execution results with rich media visualizations. However, Jupyter notebooks' linear structure and global namespace lack the module system crucial for complex software development, making them less suitable for large-scale projects and more for demonstration or visualization purposes.

Weber et al. \cite{weber2024extending} proposed extending Jupyter with multi-paradigm editors, including a graphical programming environment. Their work demonstrates the potential of combining different programming paradigms within a single environment to enhance flexibility and usability. While their focus is on multi-paradigm editing, CodePod concentrates on hierarchical organization and scoped execution to manage large projects effectively within a notebook-like setting.

The work along the line of reactive notebooks~\cite{je2021reactive} concerns the execution order of notebooks. Specifically, researchers propose analyzing the dependencies of code blocks and auto re-evaluating code blocks when the dependent code block changes. These works can improve the determinism of Jupyter notebooks and mitigate the issues of out-of-order execution and the problems that Jupyter code is prone to errors of un-executed outdated code. This work is orthogonal and complementary to CodePod. It is possible to integrate such a reactive evaluation model in CodePod to re-evaluate code blocks when the dependent code blocks change using the def-use dependencies in CodePod.

\subsection{Spatial Organization of Code}
Several research efforts have explored spatial organization to improve code navigation and understanding. Bragdon et al. \cite{bragdon2010code} introduced Code Bubbles, which uses a 2D canvas to arrange lightweight, editable code fragments. This approach allows developers to create concurrent working sets of code, improving navigation and understanding of complex codebases. CodePod adopts a similar spatial organization but extends it with hierarchical modules and scoped execution, specifically tailored for notebook-based interactive development.

Deline and Rowan \cite{deline2010code} presented Code Canvas, an infinite zoomable surface for software development that leverages spatial memory to help developers stay oriented and synthesize information. CodePod's 2D canvas serves a similar purpose but is integrated within a notebook environment, emphasizing hierarchical structure and scoped execution for large-scale projects.

Debugger Canvas \cite{deline2012debugger} is an industrial application of the code bubbles paradigm, integrated into Visual Studio for debugging. While it shares the spatial organization concept, its focus is on debugging, whereas CodePod is designed for general interactive development and project management in notebooks.

Code Canvas and Code Bubbles address navigation problems of mainstream file-based IDEs, while CodePod extends these ideas to the notebook environment, providing a scalable solution for interactive development at scale.

Kučečka et al. \cite{kuvcevcka2022uml} explored a UML-based live programming environment in virtual reality, providing an immersive experience for modeling and understanding software systems. Although this is in a different medium, it aligns with the goal of improving code comprehension through innovative interfaces. CodePod, however, focuses on a 2D canvas and hierarchical structures within a browser-based environment, ensuring compatibility with existing notebook workflows.

Code analyzers like Unified Modeling Language (UML)\cite{booch1996unified,medvidovic2002modeling} and call graph visualizers\cite{callahan1990constructing} generate visual presentations of code. However, these are non-editable and non-interactive, therefore less useful during the development process.

\subsection{Program Comprehension and Managing Complexity}
Olivero et al. \cite{olivero2011enabling} introduced Gaucho, a visual object-focused development environment that uses intuitive depictions of object-oriented constructs to aid program comprehension.
Head et al. \cite{head2019managing} addressed the issue of managing clutter and inconsistency in computational notebooks through code gathering tools.




\subsection{General Development Environments}

Glamorous Toolkit \cite{gtoolkit} is a comprehensive development environment that emphasizes making systems explainable through custom tools. While it provides a broad range of features for understanding software systems, it does not provide hierarchical scope and namespace management. CodePod's specific contribution is its hierarchical structure and scoped execution within a notebook-like interface, making it particularly suited for large-scale interactive development in data science and similar fields.



\section{Conclusion}

We have presented CodePod, a hierarchical extension of Jupyter that introduces a novel scoped execution model with formal semantics. Our work develops a mathematical framework for hierarchical scoping, enabling precise reasoning about symbol visibility and accessibility, and implements a language-agnostic runtime system that realizes these semantics through source-level transformations—supporting true incremental evaluation across nested modules without kernel modifications. This approach allows CodePod to scale interactive development both within and beyond traditional notebook boundaries. Through theoretical analysis, qualitative user studies, and large-scale validation, we demonstrate significant improvements in project scalability and developer productivity, with participants reporting substantial benefits in managing larger projects compared to Jupyter.





\bibliographystyle{ACM-Reference-Format}
\bibliography{main}


\end{document}